\documentclass[runningheads]{llncs}

\usepackage{graphicx}
\usepackage{amsmath}
\usepackage{amsfonts}
\usepackage{amssymb}
\usepackage{hyperref}
\usepackage{dsfont}

\DeclareMathOperator*{\argmin}{arg\,min}

\begin{document}

\title{Optimal Publishing Strategies on a Base Layer}

\author{Yogev Bar-On\inst{1} \and Yishay Mansour\inst{1,2}}
\authorrunning{Bar-On \& Mansour}

\institute{Tel Aviv University \and
Google Research\\
\email{\{baronyogev,mansour.yishay\}@gmail.com}}

\maketitle

\begin{abstract}
A growing number of products use layer 2 solutions to expand the capabilities of primary blockchains like Ethereum, where computation is off-loaded from the root chain, and the results are published to it in bulk. Those include optimistic and zero-knowledge rollups, information oracles, and app-specific chains. This work presents an analysis of layer 2 blockchain strategies determining the optimal times for publishing transactions on the root chain. There is a trade-off between waiting for a better layer 1 gas price and the urgency to finalize layer 2 transactions. We present a model for the problem that captures this trade-off, generalizing previous works, and we analyze the properties of optimal publishing strategies. We show that such optimal strategies hold a computable simple form for a large class of cost functions.

\end{abstract}

\section{Introduction}\label{sec:intro}
Layer 2 (L2) blockchains are designed to scale and upgrade primary and secure blockchain networks such as Ethereum \cite{wood2014ethereum} and Bitcoin \cite{nakamoto2008bitcoin} (often called the \emph{base layer}, \emph{root chain} or \emph{layer 1}). They operate by off-loading computation from the base layer, enabling faster transaction times, lower fees, higher volume, and extra functionality. By publishing on the root chain, L2 chains extend the existing root chain (almost) without sacrificing security and availability of information.

The most common are rollup chains that are designed to increase transaction throughput while decreasing their cost by batching many transactions into a single compressed transaction that can be efficiently published on the base layer. Examples include optimistic rollups like Optimism \cite{optimism2023} and Arbitrum \cite{kalodner2018arbitrum} and, more recently, zero-knowledge rollups such as Starknet and zkSync \cite{zksync2023}.

Another example is networks that provide functionality that does not exist on the root chain, such as oracles. Oracles such as Chainlink \cite{breidenbach2021chainlink}, are designed to securely provide the base layer with off-chain data, like token pricing or weather data, which normally is not accessible from base layer applications.

In all those cases, there is a trade-off between the price of publishing on the root chain and the cost of delaying publication. The cost of publishing on secure root chains (\emph{gas price}) varies and depends on network congestion and transaction complexity. On the other hand, waiting a long time for the price to be low may also be costly. In rollups, for example, transactions should be treated as pending until posted on the root chain since the security and reliability of the L2 infrastructure are less trusted. Hence there is an explicit cost of waiting for transaction finalization and a more implicit cost of a reliability issue with the rollup. For oracles, a delay can cause on-chain information to be outdated, which can have devastating effects on, e.g., pricing oracles.

\subsection{Our Contributions}
This paper proposes a general model for the decision problem of deciding when to publish L2 information on the root chain. This model encapsulates both the cost of publishing, which varies with the L1 gas price, and the cost of waiting (which we call the delay cost). We make a few simplifying assumptions that, while potentially limiting the practical applicability of the results in their current form, offer valuable insights into the fundamental methodologies that are useful in practice.

We then give a solution for the optimal strategies for a couple of classes of cost functions.

For \emph{constant publishing cost functions}, which do not depend on the amount of published information, we show that a greedy policy that tries to balance between the delay and publishing cost is near-optimal.

For \emph{homogeneous publishing cost functions}, which do not have a fixed overhead cost, we prove the optimal strategy is simply a threshold policy - meaning we publish only below a certain gas price.

We simulate our policies and validate they hold well empirically as well.

\subsection{Related Works}
To the best of our knowledge, the only prior work on the publishing time decision problem is \cite{mamageishvili2023efficient}. In their work, the authors looked at a specific cost function with quadratic delay cost suitable for optimistic rollups. They used a Q-learning method to empirically observe the optimal MDP strategy properties and compared it with other strategies used for the Arbitrum blockchain.

Our model generalizes this work, encapsulating more cost functions suitable for a wider array of layer 2 solutions, like oracles. Moreover, we give a closed-form solution for their model, and our theory proves properties empirically observed by their research. 

Rollup chains are surveyed in \cite{thibault2022blockchain}, and some prominent layer 2 blockchains include \cite{kalodner2018arbitrum,kanani2021matic,breidenbach2021chainlink,optimism2023,zksync2023}.

The L2 publication problem can also be considered as an expansion on the long-studied inventory policy problem \cite{arrow1951optimal,veinott1965optimal,huggins2010inventory,goldberg2021survey}, where the inventory demand is dual to the publishing cost. We, however, do not assume a specific form for the delay cost, where most of these works use a constant delay, with some considering a linear delay (albeit a deterministic demand) \cite{kumar2019inventory,kumar2019inventory2,tripathi2018establishment}. Similarities to our work can be found in \cite{feinberg2005optimality} which shows the optimality of a threshold policy in certain cases, and \cite{xin2022distributionally} which considers a martingale demand. 

When discussing a constant publication cost, there are similarities with sequential buy-or-rent problems (also known as ski rental problems) \cite{karlin1994competitive,gollapudi2019online,shah2021sequential}, where an agent needs to decide how long to rent each resource before buying it - similar to deciding how long to hold a transaction before publishing it. However, a key difference is that in our problem, publishing does not stop paying for resources since new transactions keep coming.

Also relevant is the dynamics of L1 gas prices, explored in \cite{roughgarden2020transaction,leonardos2021dynamical,liu2022empirical}, which analyzes gas prices on Ethereum after the introduction of EIP-1559 \cite{buterin2019eip1559}.

\subsection{Outline}
We start in the next section by giving a formulation for the L2 publishing problem and presenting the notation and assumptions we will use. In Section \ref{sec:constant}, we present a near-optimal strategy for constant publishing costs. We then classify the optimal strategy for homogeneous cost publishing costs in Section \ref{sec:homogeneous}. Section \ref{sec:rollup} applies the analysis on a specific problem instance representing optimistic rollups, justifying why our formulation generalizes previous formulations. We show simulations in Section \ref{sec:simulations} and conclude in Section \ref{sec:conclusion}. All proofs are deferred to Appendix \ref{sec:proofs} for clarity.

\section{Preliminaries}\label{sec:pre}
We consider an infinite horizon decision problem. At each time step $t$, a new transaction $H_t$ is created that encapsulates L2 information to be published on the base layer. This can be an actual transaction or some aggregated information encapsulating a single atomic update to the base layer. An agent that is responsible for publishing needs to decide if to publish the transaction given the current gas price $P_t$ of the root chain or wait for a future time step and incur a delay cost. Note that in practice the agent might not know the next effective gas price, but can only estimate it.

We model the problem as a Markov Decision Process (MDP). At time $t$, the state is $\left(t,P_t,Q_t\right)$, for some gas price $P_t$ of the base layer and a set $Q_t$ of unpublished transactions. The agent needs to pick a subset $N^\pi_t\triangleq \pi(t,Q_t,P_t)\subseteq Q_t$ of transactions to publish using policy $\pi$, incurring a non-negative cost of
\[C_t(P_t,Q_t,N^\pi_t) \triangleq P_t C_t^{(\textrm{p})}(N^\pi_t) + C_t^{(\textrm{d})}(Q_t \setminus N^\pi_t)\]
where $C_t^{(\textrm{p})}$ is the publishing cost (in terms of gas) and $C_t^{(\textrm{d})}$ is the delay cost. The next state is updated with a new transaction $Q^\pi_{t+1}=(Q_t \setminus N^\pi_t) \cup \{H_t\}$, and with the next base layer fee $P_{t+1} = R\left(P_t\right)$, where $R$ is a random function that models the fee fluctuations. The goal of the agent is to find a policy that minimizes the expected cost:
\begin{definition}\label{def:optimal_policy}
Policy $\pi^*$ is \emph{optimal} if $C(\pi^*) = \min_\pi C(\pi)$, where $C(\pi)$ is the expected total cost of policy $\pi$. In particular, for the publishing problem, we have:
\[ C(\pi) \triangleq \mathbb{E}\left[\sum_t \gamma^t C_t(P_t,Q^{\pi}_t,N^{\pi}_t)\right] \]
using a discount factor $0<\gamma<1$.
\end{definition}

During this work, we assume the agent knows the cost function, including the distribution of $R$.

\subsection{Cost}
\paragraph{Publishing cost} The cost of publishing a single transaction is usually constant in terms of the amount of gas needed, so we can assume that $C_t^{(\textrm{p})}(N)$ is linear in $|N|$ (where $N$ is the set of published transactions), and we will define
\[ C_t^{(\textrm{p})}(N)\triangleq \alpha|N|+\beta \mathds{1}[N\neq\emptyset] \]
for some parameters $\alpha,\beta\geq 0$. Note that in the case where $|N|=0$, we do not incur any publishing cost since no interaction with the base layer is needed.

\paragraph{Delay cost} For the delay cost, we assume each transaction incurs a cost independently, using its own cost function. Thus, we define
\[ C_t^{(\textrm{d})}(N)\triangleq\sum_{H_\tau \in N}{C^{(\textrm{d})}_{H_\tau}(t-\tau)} \]
where $C^{(\textrm{d})}_{H_\tau}$ is the independent delay cost of $H_\tau$ as a non-negative function of the waiting time.

In some cases, we assume a \emph{global delay cost} that does not depend on the specific transaction, i.e., $C^{(\textrm{d})}_{H_\tau} = C^{(\textrm{d})}_{H_\mu}$ for all $\tau,\mu$. In such cases, we remove the transaction from the notation and denote the delay cost by $C^{(\textrm{d})}$.

\subsubsection{Examples}
To illustrate the flexibility of this model, we will see a few different possible cost functions that can be seen as idealized models for different scenarios.

\paragraph{Optimistic rollups} Optimistic rollups gather transaction batches that are finalized on L1. Each batch is published separately and has the same urgency. This can be modeled by using a linear delay cost with coefficient $c$ and a homogeneous publishing cost:
\begin{align*}
    C_t^{(\textrm{p})}(N) = \alpha |N| && C^{(\textrm{d})}(i) = ci.
\end{align*}

\paragraph{Zero-knowledge rollups} Somewhat similar to optimistic rollups, ZK rollups also gather transaction batches. The core difference is that only a short ZK validity proof needs to be published on L1, and thus, we can use a constant publishing cost.
\[ C_t^{(\textrm{p})}(N) = \beta\mathds{1}[N\neq\emptyset]. \]

\paragraph{Oracles} Oracles are specifically designed to publish certain information on L1 (like token prices), and this information needs to be updated frequently. Hence, the delay cost may be exponential, with different possible rates $c_{H_\tau}$ for each transaction (more volatile prices must be updated more frequently). The publishing cost depends on the specific implementation, so we will leave it in the generic form.
\[ C^{(\textrm{d})}_{H_\tau}(i) = \exp\left(i~c_{H_\tau}\right). \]

\section{Constant Publishing Cost}\label{sec:constant}
We start by looking into the case where the publishing cost is constant, i.e., $\alpha=0$ and $C^{(\textrm{p})}_t(N)=\beta\mathds{1}[N\neq\emptyset]$. This is common in cases where the base layer only contains short proofs of the information from layer 2, such as zero-knowledge proofs for ZK rollups, as previously discussed. Another interesting use-case is Merkle tree roots for proof of deposits \cite{merkle1987digital}, used to bridge funds between the layers.

An immediate observation is that, in this case, there is no reason to publish only some of the transactions - since publishing all of them will cost the same.

\begin{lemma}\label{lem:all_or_nothing}
    If $C^{(\textrm{p})}_t(N)=\beta\mathds{1}[N\neq\emptyset]$, there is an optimal policy $\pi^*$ that always publishes all available transactions or none of them, i.e., either $N^{\pi^*}_t=Q^{\pi^*}_t$ or $N^{\pi^*}_t=\emptyset$.
\end{lemma}

In the following, we assume a global delay cost $C^{(\textrm{d})}$. Using Lemma \ref{lem:all_or_nothing}, we get that the amount of non-published transactions at each step is the number of steps since the last publication.

Hence, we get that the total delay cost in the interval between two publications depends only on the interval's length. This leads us to the following definition:
\begin{definition}\label{def:agg_cost}
    In the case of a global delay and a constant publishing cost, we define the \emph{aggregated delay cost} $F^{(\textrm{d})}$ as:
    \[ F^{(\textrm{d})}(n) \triangleq \sum_{t=1}^{n-1}\sum_{i=1}^t \gamma^{t-1} C^{(\textrm{d})}(i) \]
\end{definition}
Using this definition, the total delay cost between two publications is $\gamma^t F^{(\textrm{d})}(n)$ where $t$ is the first step of the interval, and $n$ is the length of the interval.

\subsection{Constant Gas Price}
To gain more intuition on the problem, we will start with the case where the prices are constant, i.e., $R(P)=P$ with probability $1$. In this case, the optimal policy is to publish in constant intervals, as we prove in the following theorem.

\begin{theorem}\label{thm:const_price}
    If the delay cost is global, the gas price is constant and equals to $P$, and the publishing cost is constant and equals to $\beta$, an optimal policy is to publish in constant intervals of size $n^*$, chosen to minimize the average cost of a single interval:  
    \[ n^* = \argmin_n \frac{F^{(\textrm{d})}(n) + \gamma^{n - 1} \beta P}{1-\gamma^n}.\]
\end{theorem}

To see how we can use Theorem \ref{thm:const_price} to obtain a specific solution, we can use a linear delay cost $C^{(\textrm{d})}(i)=6i$ and approximate $\gamma\approx 1$ for simplicity. Using this delay function, the aggregated delay cost becomes:
\[ F^{(\textrm{d})}(n) = \sum_{t=1}^{n-1}\sum_{i=1}^t 6i = \sum_{t=1}^{n-1} 3t(t+1) = n(n^2-1), \]
and thus we get:
\[ n^* = \argmin_n \frac{n(n^2-1) + \gamma^{n - 1} \beta P}{1-\gamma^n}.  \]
When taking $\gamma\rightarrow 1$, the minimum is obtained over the reals for $n = \sqrt[3]{\frac{1}{2}\beta P}$ (can be shown by differentiating), so we get $n^*$ by rounding that number to a near integer.

We notice that if we assign the minimal value to the cost, we get that the delay cost is $n^* ({n^*}^2-1)=\frac{\beta P}{2} - \sqrt[3]{\frac{1}{2}\beta P}$, in the same order of magnitude as the publishing cost.

\subsection{Near-Optimal Policy}
Trying to generalize Theorem \ref{thm:const_price}, one can hypothesize that it is a common case that an optimal policy has about the same delay and publishing cost, even if the gas price is not constant. In the following, we show that for certain delay costs, it is indeed near-optimal.

We start by defining the class of delay costs for which our strategy works:
\begin{definition}\label{def:sub_add}
    An aggregated cost function $F^{(\textrm{d})}$ is said to be \emph{$\sigma$-sub-additive} if for all $n_1,n_2\geq 1$:
    \[ F^{(\textrm{d})}(n_1 + n_2) \leq \sigma\left(F^{(\textrm{d})}(n_1+1) + \gamma^{n_1} F^{(\textrm{d})}(n_2)\right). \]
\end{definition}
Intuitively, it means that adding a publishing step does not reduce the delay cost too much (more than $\sigma$). As an example, consider the aggregated delay cost we used previously (ignoring $\gamma$): $F^{(\textrm{d})}(n)=n(n^2-1)$. It is easy to verify this function is $4$-sub-additive.

To keep the delay and publishing costs similar, we suggest the following strategy: keep track of the amount of delay cost incurred since the last publication (this can be computed from the number of available transactions) and publish when the current publication cost is lower.

\begin{theorem}\label{thm:near_optimal}
    Assume the delay cost is global such that the aggregated cost function is $\frac{\sigma}{2}$-sub-additive, and the publishing cost is constant and equals $\beta$. Let $\pi$ be the policy as described above:
    \[ \pi(t,Q_t,P_t) = \begin{cases}
    Q_t  & \gamma^{|Q_t|-1} \beta P_t \leq F^{(\textrm{d})}(|Q_t|+1) \\
    \emptyset & \text{else}
    \end{cases}. \]
    Then $\pi$ is $\sigma$-optimal, i.e., $C(\pi)\leq \sigma C(\pi^*)$ where $\pi^*$ is an optimal policy.
\end{theorem}
It is important to note that we did not assume how gas prices are distributed, making this result useful for many cases.

\section{Homogeneous Publishing Cost}\label{sec:homogeneous}
We now consider the case where the publishing cost is homogeneous, i.e., $\beta=0$ and $C^{(\textrm{p})}_t(N)=\alpha|N|$. We notice that, in this case, we can treat each transaction independently.

Consider an independent MDP for transaction $H_\tau$ with state $\left(t,P\right)$ at time $t$ and price $P$.  We define the independent publishing cost function of $H_\tau$ to be:
\[ C^{(\textrm{p})}_{t,H_\tau}(P) \triangleq \alpha P, \]
and the independent delay cost to be:
\[ C^{(\textrm{d})}_{t,H_\tau}(P) \triangleq C^{(\textrm{d})}_{H_\tau}(t-\tau). \]

At each step, the agent chooses whether to publish the transaction or wait.
\begin{definition}\label{def:independent_policy}
    An \emph{independent policy} $\pi_{H_\tau}(t,P)\subseteq \{H_\tau\}$ for transaction $H_\tau$ decides if to publish the transaction at step $t$ with price $P$ ($\pi_{H_\tau}(t,P)=\{H_\tau\}$) and incur cost $C^{(\textrm{p})}_{t,H_\tau}(P)$ or wait ($\pi_{H_\tau}(t,P)=\emptyset$) and incur cost $C^{(\textrm{d})}_{t,H_\tau}(P)$.
\end{definition}

Note that for transaction sets $N,Q$:
\begin{align*}
&\sum_{H_\tau \in N} C^{(\textrm{p})}_{t,H_\tau}(P) + \sum_{H_\tau \in Q \setminus N} C^{(\textrm{d})}_{t,H_\tau}(P) \\
&= \alpha |N| P + \sum_{H_\tau \in Q \setminus N} C^{(\textrm{d})}_{H_\tau}(t-\tau) \\
&= C_t(P,Q,N),
\end{align*}
and thus, finding the optimal strategy for each transaction independently is equivalent to finding the optimal strategy for all transactions, summarized in the following:
\begin{lemma}\label{lem:independent}
Let $\pi^*_{H_\tau}(t,P)$ be an optimal independent policy for transaction $H_\tau$. Then:
\[ \pi^*(t,P,Q) = \bigcup_{H_\tau \in Q}{\pi^*_{H_\tau}(t,P)} \]
is an optimal policy for the publishing problem with a homogeneous publishing cost.
\end{lemma}

\subsection{Price Threshold}
The optimal independent policy takes a relatively simple form for a certain class of price fluctuation distributions called \emph{monotonically non-expansive}.
\begin{definition}\label{def:contracting}
    A random function $R(P)$ is \emph{monotonically non-expansive} if $\mathbb{E}\left[R(P)\right]$ is both non-expansive and monotonically increasing, i.e.,  for all $P_2\geq P_1$:
    \[ 0 \leq \mathbb{E}\left[R(P_2)\right] - \mathbb{E}\left[R(P_1)\right] \leq P_2-P_1.  \]
\end{definition}
The requirement for monotonicity is fairly intuitive since, as the current price is higher, we expect the next price to be higher as well. The non-expansion requirement can mean that either the prices change as a random walk or that there is some fixed price that the prices gradually move towards (due to the Banach fixed-point theorem). Both are common assumptions (see, e.g., \cite{metcalf1995investment}), and as we will see, they also hold for Ethereum gas prices.

We can now provide a classification of optimal independent policies.
\begin{theorem}\label{thm:independent_threshold}
 If the price fluctuation function $R$ is monotonically non-expansive, there exists a monotonically increasing price threshold $\lambda_{H_\tau}(x)$ such that:
\[
\pi^*_{H_\tau}(t,P) = \begin{cases}
\{H_\tau\}  & P \leq \lambda_{H_\tau}(t-\tau) \\
\emptyset & P > \lambda_{H_\tau}(t-\tau) 
\end{cases}
\]
is an optimal independent policy for transaction $H_\tau$. Moreover, any two transactions with the same delay cost will have the same price threshold.
\end{theorem}

Hence, an optimal policy is simply to check if the price is above or below a certain threshold that depends on the specific instance parameters and the amount of time a transaction is waiting.

Using Lemma \ref{lem:independent}, we can also classify the general problem.
\begin{corollary}\label{cor:threshold}
If the price fluctuation function $R$ is monotonically non-expansive, there exists monotonically increasing price thresholds $\lambda_{H_\tau}(x)$ for any transaction $H_\tau$ such that:
\[
\pi^*(t,P,Q) = \left\{H_\tau \mid H_\tau \in Q \land P \leq \lambda_{H_\tau}(t-\tau)\right\}
\]
is an optimal policy for a homogeneous publishing cost.
\end{corollary}

This result is a theoretical confirmation of the empirical observations made in \cite{mamageishvili2023efficient}, where $\min_{H_\tau \in Q}\lambda_{H_\tau}(t-\tau)$ is the price threshold for publishing all transactions, and $\max_{H_\tau \in Q}\lambda_{H_\tau}(t-\tau)$ is the price threshold for not publishing any transaction.

\section{Rollup Cost Function}\label{sec:rollup}

We now explore the case where the delay cost function is global and monotonic. Intuitively, in this case, there is no reason to publish a new transaction before an old one, and we expect the optimal algorithm to publish transactions by order of arrival. Hence, we can simplify our MDP such that the agent only needs to choose the number of transactions to publish rather than which transactions to publish (given the amount, it is always the earliest ones).

We start by formalizing and proving the intuitive claim.
\begin{lemma}\label{lem:monotone}
If the delay cost function is global and monotonically increasing, the earliest transactions are published when using an optimal policy $\pi^*$. In other words, for all $H_\tau\in N^{\pi^*}_t$ and $H_\nu\in Q^{\pi^*}_t \setminus N^{\pi^*}_t$, we have $\tau<\nu$.
\end{lemma}

Let us now define a modified MDP with the state at time $t$ being $(t,P_t,\hat{Q}_t)$ where $\hat{Q}_t$ denotes the \emph{number} of unpublished transactions at time $t$. The agent only needs to choose $\hat{N}^{\hat{\pi}}_t\in [0,\hat{Q}_t]$ and the state updates with $\hat{Q}^{\hat{\pi}}_{t+1}=\hat{Q}_t-\hat{N}^{\hat{\pi}}_t+1$ (the price changes as before). The agent then incurs a cost of:
\[ \hat{C}_t(P_t,\hat{Q}_t,\hat{N}^{\hat{\pi}}_t) \triangleq P_t (\alpha \hat{N}^{\hat{\pi}}_t+\beta) + \sum_{i=1}^{\hat{Q}_t-\hat{N}^{\hat{\pi}}_t} C^{(\textrm{d})}(i). \]

Using Lemma \ref{lem:monotone}, it's easy to see the modification does not change the optimal strategy:
\begin{corollary}\label{cor:equiv_mdp}
    Let $\pi^*$ be an optimal policy for the original MDP. Then
    \[ \hat{\pi}^*(t,P_t,|Q^{\pi^*}_t|) = |\pi^*(t,P_t,Q^{\pi^*}_t)| \]
    is an optimal policy for the modified MDP.
\end{corollary}

Hence, a model with the following parameters:
\[C^{(\textrm{d})}(i)=2ci,\alpha=1,\beta=0\]
for some unit conversion factor $c>0$ (which implies a homogeneous publishing cost), is equivalent to the simplified model with
\begin{align*}
  \hat{C}_t(P,\hat{Q},\hat{N}) 
  &= P \hat{N} + \sum_{i=1}^{\hat{Q}-\hat{N}} 2ci \\
  &= P \hat{N} + c(\hat{Q}-\hat{N})(\hat{Q}-\hat{N}+1).
\end{align*}
This is the same quadratic model presented in \cite{mamageishvili2023efficient} for rollup chains. This can be justified by thinking of the risk in rollup chains to be malicious actions due to unpublished transactions. These actions may occur with a small, constant probability at each time step and thus result in a delay cost that increases linearly over time.
We can use our analysis to solve this specific problem instance.

\subsection{Price Dynamics}
We want to model $R$ similar to how the Ethereum base fee behaves. Since EIP-1559, each block changes the base fee depending on transaction load and multiplies the base fee by a factor between $\frac{7}{8}$ and $\frac{9}{8}$. We let $X_i \in \left[\frac{7}{8},\frac{9}{8}\right]$ be a random variable that models the base fee factor after block $i$. Assuming $k$ blocks are produced between two time steps, we get $P_{t+1} = P_t \prod_{i=1}^k X_i$. Assuming the simple (although not realistic) case where $X_i$'s are i.i.d, for a large enough $k$ we can approximate $\prod_{i=1}^k X_i$ as a log-normal distribution due to the Central Limit Theorem. Hence, we will define:
\[ R(P) \triangleq Pe^{\mathcal{N}(\mu,\sigma^2)} \]
where $\mathcal{N}(\mu,\sigma^2)$ is a normal distribution with mean $\mu$ and variance $\sigma^2$.

We have $\mathbb{E}\left[e^{\mathcal{N}(\mu,\sigma^2)}\right]=e^{\mu+\frac{\sigma^2}{2}}$, so for $\mu\leq -\frac{\sigma^2}{2}$ we get that $R$ is monotonically non-expansive.

\begin{figure}[htbp]
    \centering
    \includegraphics[width=0.9\textwidth]{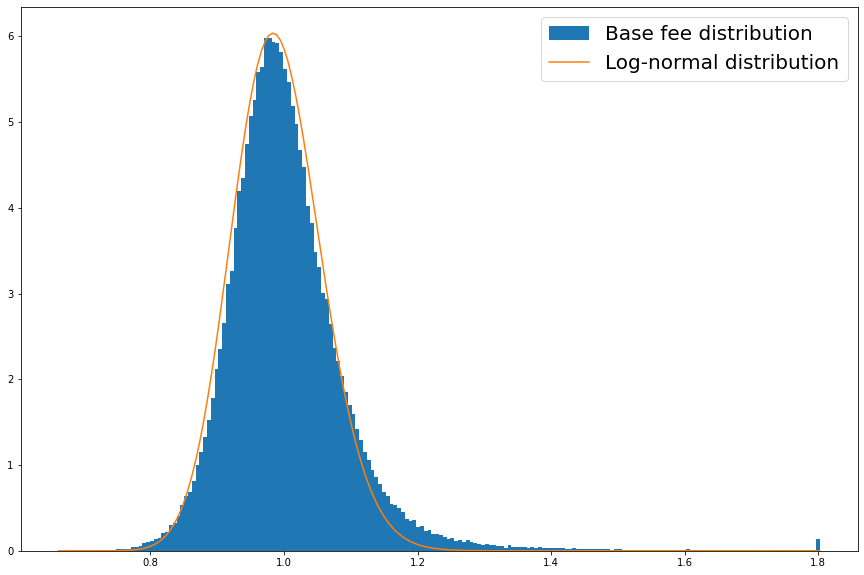}
    \caption{Empirical minutely Ethereum base-fee factor distribution, taken from block 15M to 17M, next to a log-normal distribution.}
    \label{fig:fee_dist}
\end{figure}

In Figure \ref{fig:fee_dist}, we show the empirical Ethereum minutely base fee factor distribution and compare it to a log-normal distribution, making the case they are very similar in practice. Note there is a small outlier for large factors, which can be explained in some theoretical setups \cite{hougaard2023farsighted}.

\subsection{Optimal Policy}
Since the rollup model has homogeneous publishing cost and monotonically non-expansive price fluctuations (when $\mu \leq -\frac{\sigma^2}{2}$), we can use Corollary \ref{cor:threshold} and get the optimal policy by finding the price threshold.

\begin{theorem}\label{thm:optimal_threshold}
If $\mu \leq -\frac{\sigma^2}{2}$, using
\begin{equation}\label{eq:optimal_threshold}
\lambda(x)=\frac{2c}{1-\gamma} \inf_{n\geq 1} \frac{(1-\gamma^n)(x+\frac{\gamma}{1-\gamma})-n\gamma^n}{1-\gamma^n e^{n(\mu+\frac{\sigma^2}{2})}}
\end{equation}
as a price threshold for all transactions is an optimal strategy for the rollup model.
\end{theorem}

We can interpret Eq. \ref{eq:optimal_threshold} by considering $n$ to be the number of future time steps we wait before publishing the transaction. The price threshold is then the best total cost possible by waiting in the current step. It can be easily computed using numeric methods, or one could choose to approximate it optimistically by using $n=1$:
\[ \lambda(x) \leq \frac{2cx}{1-\gamma e^{\mu+\frac{\sigma^2}{2}}}. \]
Using this approximation, the policy considers waiting a single step as the only alternative to publishing.

Note the special case $\mu=-\frac{\sigma^2}{2}$ that implies that $R$ is a martingale process, i.e., $\mathbb{E}\left[R(P)\right]=P$; meaning the expected future price is the current price.

We can get an analytical solution for the price threshold using Theorem \ref{thm:optimal_threshold}.
\begin{corollary}\label{cor:martingale}
If $\mu=-\frac{\sigma^2}{2}$, using
\[ \lambda(x) = \frac{2cx}{1-\gamma} \]
as a price threshold is an optimal strategy for the rollup model.
\end{corollary}

\section{Simulations}\label{sec:simulations}
To see how our strategies hold in practice, we ran simulations and compared the results to the trivial strategy of publishing at every step. In all experiments we used $\gamma=1-10^{-7}$ and a global delay cost of $C^{(\textrm{d})}(i)=(1-\gamma)i$.

\subsection{Homogeneous Publishing Cost}
We start with simulating the case where $\alpha=1$ and $\beta=0$ as in the rollup case. Assuming the gas price dynamic is martingale, we use the price threshold $\lambda(x) = x$, which, according to Corollary \ref{cor:martingale} is optimal.

\begin{figure}[htbp]
    \centering
    \includegraphics[width=0.9\textwidth]{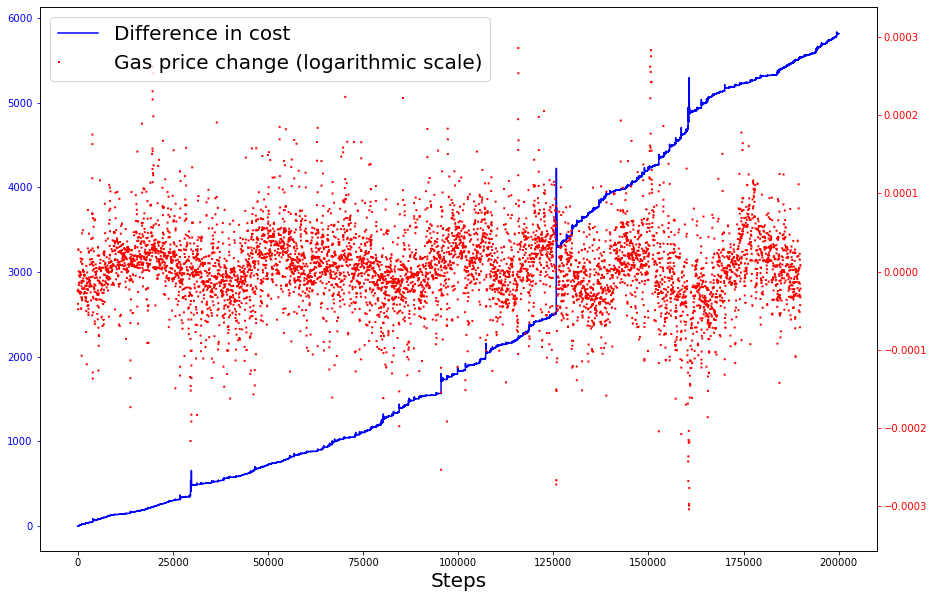}
    \caption{The difference of the total cost over time for the price threshold policy and the trivial policy, on minutely Ethereum base fees taken from block 15M to 17M.}
    \label{fig:real_diff}
\end{figure}

Figure \ref{fig:real_diff} shows the aggregated difference between the trivial and the optimal policies using historical Ethereum minute base fees over about nine months. The maximal waiting time of our strategy was 46 minutes. Notice that our policy consistently outperforms the trivial one and that in cases where the gas price is more volatile, the performance is even better. 

\begin{figure}[htbp]
    \centering
    \includegraphics[width=0.9\textwidth]{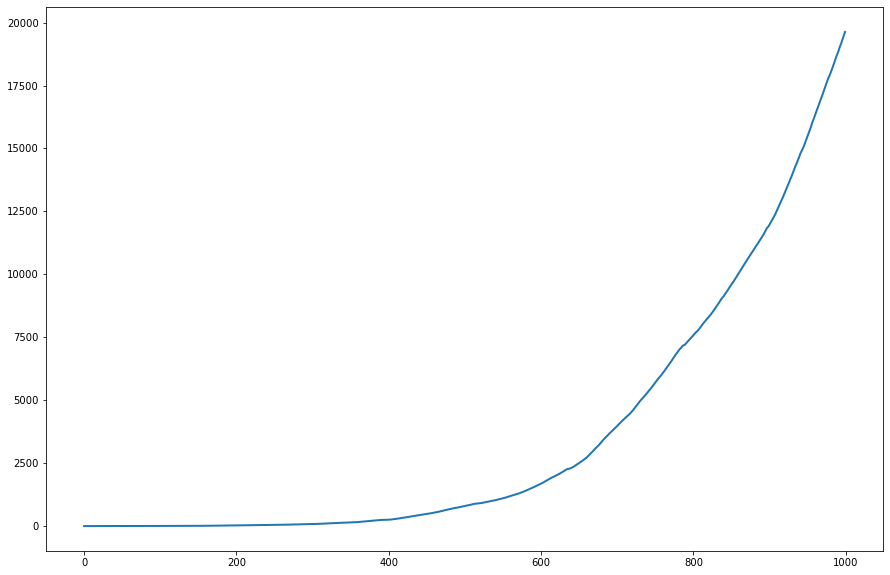}
    \caption{The mean difference of the total cost over time for the price threshold and trivial policies, using a log-normal martingale price distribution.}
    \label{fig:simulated_diff}
\end{figure}

\begin{figure}[htbp]
    \centering
    \includegraphics[width=0.9\textwidth]{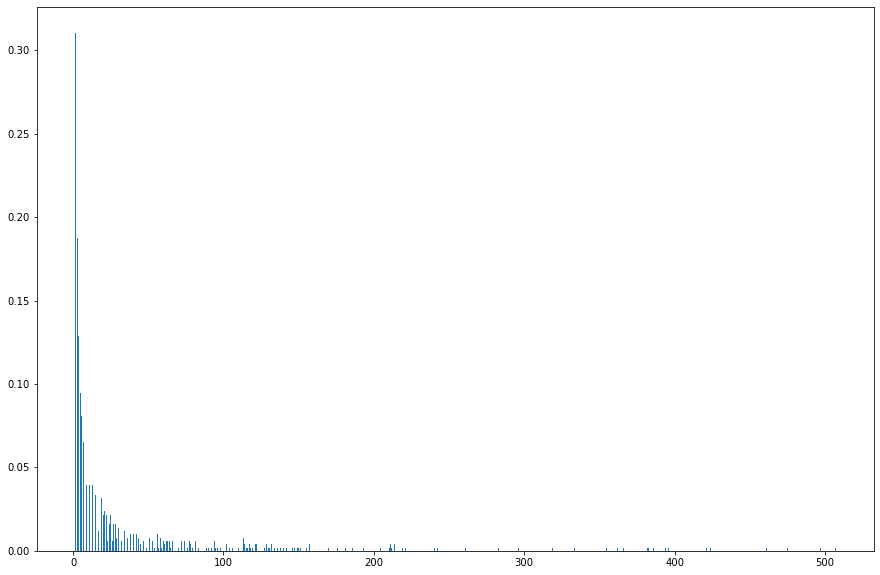}
    \caption{The distribution of the maximal waiting time for the price threshold policy using a log-normal martingale price distribution.}
    \label{fig:max_delay_hist}
\end{figure}

In Figure \ref{fig:simulated_diff}, we show the mean difference between the policies using random log-normal values (such that $\mu=\frac{\sigma^2}{2}$) over experiments with 1000 steps each. Here, we can see that the performance is much better, fitting our theory and hinting at the need to improve the Ethereum gas price distribution model. The average maximal waiting time was 35.79 steps, and we show the maximal waiting time distribution in Figure \ref{fig:max_delay_hist}. We can see that the waiting time is usually low, but there is a long tail of longer waiting times.

\subsection{Constant Publishing Cost}
We now simulate using $\alpha=0$ and $\beta=1$. We use our near-optimal policy from Theorem \ref{thm:near_optimal} and again compare it to the trivial policy.

\begin{figure}[htbp]
    \centering
    \includegraphics[width=0.9\textwidth]{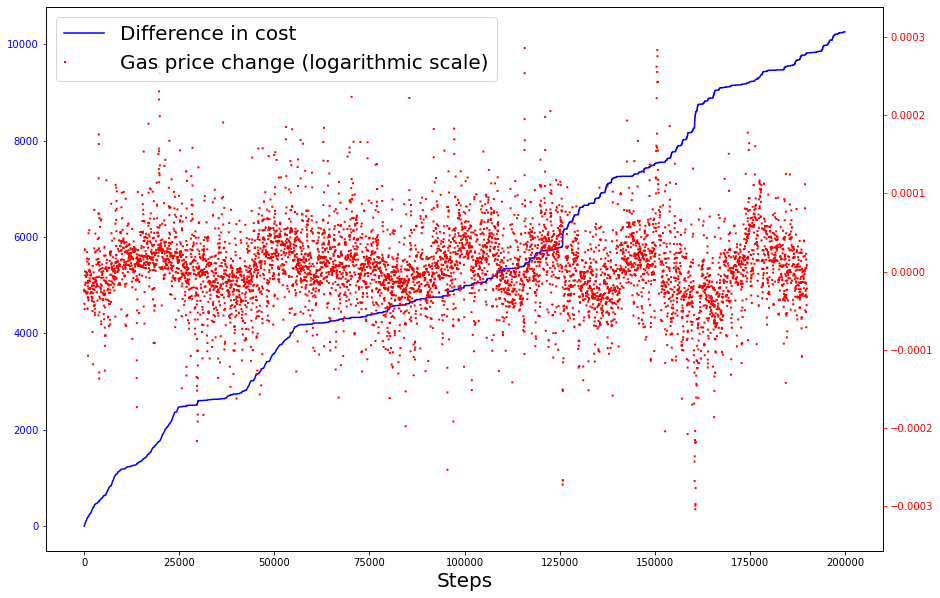}
    \caption{The difference of the total cost over time for the near-optimal constant publishing cost policy and the trivial policy, on minutely Ethereum base fees taken from block 15M to 17M.}
    \label{fig:real_diff_const}
\end{figure}

As before, we show the aggregated difference between our and trivial policies in Figure \ref{fig:real_diff_const}, using historical Ethereum minute base fees. The maximal waiting time of our strategy this time was 7 minutes. Again, our policy consistently outperforms the trivial one, but this time, the performance is more consistent and does not change drastically when the gas price is more volatile. 

\begin{figure}[htbp]
    \centering
    \includegraphics[width=0.9\textwidth]{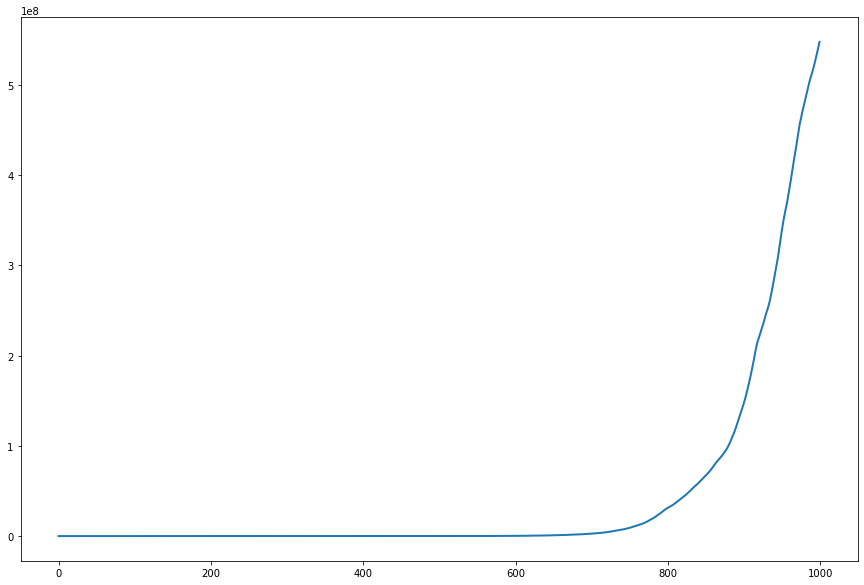}
    \caption{The mean difference of the total cost over time for the near-optimal constant publishing cost policy and the trivial policy, using a log-normal price distribution with a positive drift.}
    \label{fig:simulated_diff_const}
\end{figure}

\begin{figure}[htbp]
    \centering
    \includegraphics[width=0.9\textwidth]{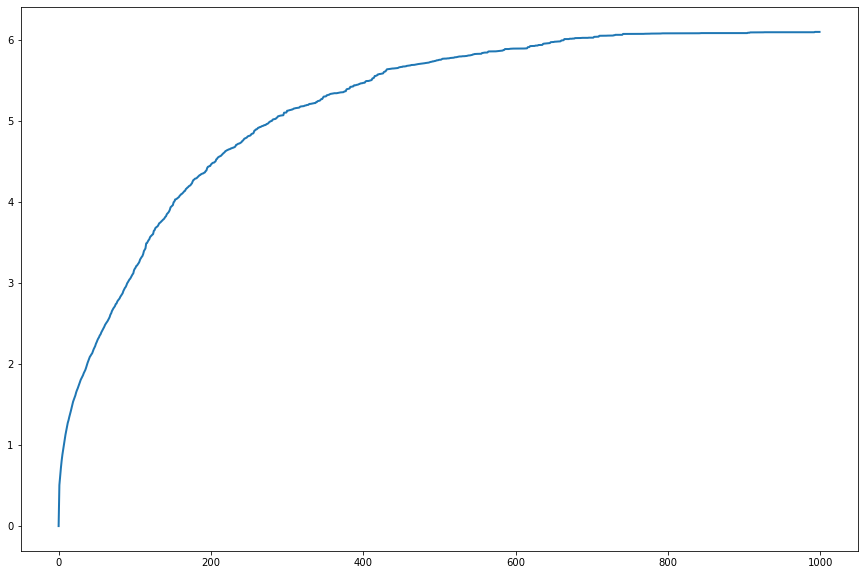}
    \caption{The mean difference of the total cost over time for the near-optimal constant publishing cost policy and the trivial policy, using a log-normal price distribution with a negative drift.}
    \label{fig:simulated_diff_const_minus}
\end{figure}

Our policy does not assume anything about the price distribution. Hence, we try it with random log-normal values both when $\mu>\frac{\sigma^2}{2}$ (Figure \ref{fig:simulated_diff_const}) and when $\mu<\frac{\sigma^2}{2}$ (Figure \ref{fig:simulated_diff_const_minus}). We show the mean difference between the policies over experiments with 1000 steps each. As expected, our policy outperforms the trivial policy in both cases. However, it is interesting to see that for the negative drift, the difference is negligible. This makes sense since, eventually, the prices are low enough such that the optimal policy is the trivial policy of publishing every step.

\section{Conclusion}\label{sec:conclusion}
We expand previous work on the decision problem of publishing transactions in the root chain, where the gas fees can vary depending on publication time. We focus on a theoretical formulation for the problem, generalizing it for many potential instances and finding the optimal strategy for a given class of parameters.

Layer 2 blockchains are becoming increasingly common, and more work needs to be done to optimize publishing strategies for more realistic scenarios. Future work may expand on our model by finding optimal policies for a larger class of cost functions, including non-homogeneous publishing cost, adversarial delay cost, and more realistic gas price fluctuation models.

\section*{Acknowledgments}
This project has received funding from the European Research Council (ERC) under the European Union’s Horizon 2020 research and innovation program (grant agreement No. 882396), by the Israel Science Foundation, the Yandex Initiative for Machine Learning at Tel Aviv University and a grant from the Tel Aviv University Center for AI and Data Science (TAD).

\bibliography{refs}

\appendix

\section{Proofs}\label{sec:proofs}
\subsection{Lemma \ref{lem:all_or_nothing}}
\begin{proof}
    Assume $\pi$ is an optimal policy such that in some step $t$ we have $\emptyset \neq N^{\pi}_t\subset Q^{\pi}_t$, and let $H_\tau \in Q^{\pi}_t \setminus N^{\pi}_t$. Define $\pi'$ to be the same policy as $\pi$ with two differences: $N^{\pi'}_t = N^{\pi}_t \cup \{H_\tau\}$ and $N^{\pi'}_{\tilde{t}} = N^{\pi}_{\tilde{t}} \setminus \{H_\tau\}$ for all $\tilde{t}>t$.
    
    Since $N^{\pi}_t \neq \emptyset$ and the publishing cost is constant, adding $H_\tau$ does not incur an extra cost. Also, removing $H_\tau$ from $N^{\pi}_{\tilde{t}}$ can only reduce the total cost since the delay cost is non-negative. Thus, $\pi'$ has a total cost at least as low as $\pi$, meaning it is optimal as well. In summary, we can always move transactions to be published in the first publication step of $\pi$ where they are available and still have optimal total cost. We thus get that publishing all the available transactions during $\pi$'s publication steps is an optimal policy.
\qed\end{proof}

\subsection{Theorem \ref{thm:const_price}}


\begin{proof}
   From Lemma \ref{lem:all_or_nothing}, we get that there is an optimal policy $\pi^*$ where all of the available transactions are published in a publication step. Hence, the policy is defined by the lengths of intervals between publications. The cost of interval $i$, which we denote by $C_{(n_1, n_2, \dots)}(i)$ for interval lengths $(n_1, n_2, \dots)$, is then:
   \[ C_{(n_1, n_2, \dots)}(i) = \gamma^{\sum_{j=1}^{i-1} n_j} (F^{(\textrm{d})}(n_i) + \gamma^{n_i - 1} \beta P) \]
   Let $(n^*_1, n^*_2, \dots)$ be the interval lengths of $\pi^*$. For it to be optimal, it is sufficient to have:
   \begin{align*}
   & (n^*_1, n^*_2, \dots) \\
   &= \argmin_{(n_1, n_2, \dots)} \sum_i C_{(n_1, n_2, \dots)}(i) \\
   &= \argmin_{(n_1, n_2, \dots)} C_{(n_1, n_2, \dots)}(1) + \gamma^{n_1} \sum_i C_{(n_2, n_3, \dots)}(i).
   \end{align*}
   Since $C_{(n^*_1, n_2, \dots)}(1)$ depends only on the first interval size, we get:
    \begin{align*}
   & (n^*_2, n^*_3, \dots) \\
   &=  \argmin_{(n_2,n_3, \dots)} C_{(n^*_1, n_2, \dots)}(1) + \gamma^{n^*_1} \sum_i C_{(n_2, n_3, \dots)}(i) \\
   &= \argmin_{(n_2,n_3, \dots)} \gamma^{n^*_1} \sum_i C_{(n_2, n_3, \dots)}(i) \\
   &= \argmin_{(n_2,n_3, \dots)} \sum_i C_{(n_2, n_3, \dots)}(i) \\
   &= (n^*_1, n^*_2, \dots).
   \end{align*}
    We thus get that all the interval sizes are the same. Let then $n^*$ be the constant interval size of $\pi^*$. We obtain:
    \begin{align*}
    &n^* \\
    &= \argmin_n \sum_i C_{(n, n, \dots)}(i) \\
    &= \argmin_n \sum_i \gamma^{n(i-1)} (F^{(\textrm{d})}(n) + \gamma^{n - 1} \beta P) \\
    &= \argmin_n~(F^{(\textrm{d})}(n) + \gamma^{n - 1} \beta P) \sum_i \gamma^{n(i-1)} \\
    &= \argmin_n \frac{F^{(\textrm{d})}(n) + \gamma^{n - 1} \beta P}{1-\gamma^n}
    \end{align*}
    as desired.
\qed\end{proof}

\subsection{Theorem \ref{thm:near_optimal}}
\begin{proof}
    Let $t_1,t_2$ be two consecutive publishing steps for $\pi$. The delay cost incurred by $\pi$ during the interval between those steps is $\gamma^{t_1} F^{(\textrm{d})}(t_2-t_1)$. Let $t^*< t_2$ be the last publishing step of $\pi^*$ before $t_2$.
    
    Since $F^{(\textrm{d})}$ is monotonically increasing, if $t^* \leq t_1$, we get that the delay cost for $\pi^*$ during this interval is larger then the delay cost for $\pi$:
    \[ \gamma^{t^*} F^{(\textrm{d})}(t_2-t^*) \geq \gamma^{t^*} F^{(\textrm{d})}(t_2-t_1) \geq \gamma^{t_1} F^{(\textrm{d})}(t_2-t_1). \]

    Otherwise, $\pi^*$ published during this interval. From the definition of $\pi$ and the fact that it did not publish in $t^*$, we get that $\pi^*$'s publishing cost was higher than $\pi$'s delay cost up to this step:
    \[ \gamma^{t^*-t_1-1}\beta P_{t^*} \geq F^{(\textrm{d})}(t^*-t_1+1) \]
    We can now use sub-additivity and show that the delay cost of $\pi$ during this interval holds:
    \begin{align*}
    & \gamma^{t_1} F^{(\textrm{d})}(t_2-t_1) \\
    &\leq \frac{\gamma^{t_1}\sigma}{2}\left( F^{(\textrm{d})}(t^*-t_1+1) + \gamma^{t^*-t_1} F^{(\textrm{d})}(t_2-t^*) \right) \\
    &\leq  \frac{\sigma}{2}\left(\gamma^{t^*-1}\beta P_{t^*} + \gamma^{t^*} F^{(\textrm{d})}(t_2-t^*)\right).
    \end{align*}

    In both cases, we get that the delay cost of $\pi$ in the interval is lower than the total cost (both delay and publishing) of $\pi^*$ in the interval, up to a factor of $\frac{\sigma}{2}$. Since the total cost of $\pi$ is at most twice the delay cost (the publishing cost is at most the delay cost), we get that $C(\pi)\leq \sigma C(\pi^*)$ as needed.
\qed\end{proof}

\subsection{Theorem \ref{thm:independent_threshold}}
\begin{proof}
An optimal independent policy $\pi^*_{H_\tau}$, by definition, publishes at state $(t,P)$ only if waiting for $n$ steps incurs an expected larger cost than publishing for all $n\geq 1$. Specifically, if for all $n\geq 1$ we have:
\begin{equation}\label{eq:publish_threshold}
\alpha P \leq \alpha \gamma^n \mathbb{E}\left[R^{[n]}(P)\right] + \sum_{i=0}^{n-1}{\gamma^i C^{(\textrm{d})}_{H_\tau}(t-\tau+i)}
\end{equation}
where $R^{[n]}$ denotes composing $R$ with itself $n$ times.

Let $P_2>P_1$. Since $R$ monotonically increases, we have that $R^{[n]}(P_2)\geq R^{[n]}(P_1)$ for all $n$. Using this fact, since $R$ is non-expansive, we get:
\[ R^{[n]}(P_2)-R^{[n]}(P_1) \leq R^{[n-1]}(P_2)-R^{[n-1]}(P_1) \leq \dots \leq P_2 - P_1. \]

Now, assume the optimal policy publishes at state $(t,P_2)$. We have that for all $n\geq 1$:
\begin{align*}
&0 \leq \alpha \left(\gamma^n \mathbb{E}\left[R^{[n]}(P_2)\right] - P_2\right) + \sum_{i=0}^{n-1}{\gamma^i C^{(\textrm{d})}_{H_\tau}(t-\tau+i)} \\
&= \alpha \left(\gamma^n \left(\mathbb{E}\left[R^{[n]}(P_2)\right] - \mathbb{E}\left[R^{[n]}(P_1)\right]\right) + P_1 - P_2\right) \\
&+ \alpha \left(\gamma^n \mathbb{E}\left[R^{[n]}(P_1)\right] - P_1\right) + \sum_{i=0}^{n-1}{\gamma^i C^{(\textrm{d})}_{H_\tau}(t-\tau+i)} \\
&\leq \alpha \left(\gamma^n \mathbb{E}\left[R^{[n]}(P_1)\right] - P_1\right) + \sum_{i=0}^{n-1}{\gamma^i C^{(\textrm{d})}_{H_\tau}(t-\tau+i)}
\end{align*}
and thus the optimal policy publishes at state $(t,P_1)$ as well.

Define $\lambda_{H_\tau}(t-\tau)$ to be the maximal price such that the optimal policy will publish the transaction at time $t$. We get that for all $P\leq \lambda_{H_\tau}(t-\tau)$ the transaction is published. Otherwise, by the definition of $\lambda_{H_\tau}$, the transaction is not published for all $P> \lambda_{H_\tau}(t-\tau)$ as desired.
\qed\end{proof}

\subsection{Lemma \ref{lem:monotone}}
\begin{proof}
Let $\pi$ be a policy such that at some step $\tilde{t}$, there exists a batch $H_\tau\in N^\pi_{\tilde{t}}$ and transaction $H_\nu\in Q^\pi_{\tilde{t}}\setminus N^\pi_{\tilde{t}}$ such that $\tau>\nu$. Also denote by $\hat{t}$ the time step where $H_\nu \in N^\pi_{\hat{t}}$. Now, let $\pi'$ a policy equal to $\pi$ except that we switch the two transactions:
\[N^{\pi'}_{\tilde{t}}=(N^\pi_{\tilde{t}} \setminus \{H_\tau\})\cup \{H_\nu\}\]
\[N^{\pi'}_{\hat{t}}=(N^\pi_{\hat{t}} \setminus \{H_\nu\})\cup \{H_\tau\}.\]

Note it means that for all $t\leq\tilde{t}$ and $t>\hat{t}$, $Q^{\pi'}_t = Q^{\pi}_t$. For all other $\tilde{t}<t\leq\hat{t}$, we have $Q^{\pi'}_t = (Q^{\pi}_t \setminus \{H_\nu\})\cup \{H_\tau\}$.

Hence $N^{\pi}_t \subseteq Q^{\pi'}_t$ for all $t\notin\{\tilde{t},\hat{t}\}$ and so $\pi'$ is properly defined. Moreover, since the actions are equal except those two times, we get $|N^{\pi}_t|=|N^{\pi'}_t|$ for all $t$.

Thus we get:
\begin{align*}
&C_t(P_t,Q^\pi_t,N^\pi_t) \\
&= P_t C_t^{(\textrm{p})}(N^\pi_t) + C_t^{(\textrm{d})}(Q^\pi_t \setminus N^\pi_t) \\
&= P_t (\alpha|N^\pi_t|+\beta) + \sum_{H_i\in Q^\pi_t \setminus N^\pi_t }C^{(\textrm{d})}(t-i) \\
&= P_t (\alpha|N^{\pi'}_t|+\beta) + \sum_{H_i\in Q^\pi_t \setminus N^\pi_t }C^{(\textrm{d})}(t-i).
\end{align*}

So for all $t<\tilde{t}$ and $t>\hat{t}$ we have:
\begin{align*}
&C_t(P_t,Q^\pi_t,N^\pi_t) \\
&= P_t (\alpha|N^{\pi'}_t|+\beta) + \sum_{H_i\in Q^{\pi'}_t \setminus N^{\pi'}_t}C^{(\textrm{d})}(t-i) \\
&= C_t(P_t,Q^{\pi'}_t,N^{\pi'}_t),
\end{align*}   

and for $\tilde{t}\leq t\leq\hat{t}$:
\begin{align*}
&C_t(P_t,Q^\pi_t,N^\pi_t) \\
&= P_t (\alpha|N^{\pi'}_t|+\beta) + \sum_{H_i\in \left((Q^{\pi'}_t \setminus \{H_\tau\})\cup \{H_\nu\}\right) \setminus N^{\pi'}_t }C^{(\textrm{d})}(t-i) \\
&= P_t (\alpha|N^{\pi'}_t|+\beta) + \sum_{H_i\in Q^{\pi'}_t \setminus N^{\pi'}_t}C^{(\textrm{d})}(t-i) - C^{(\textrm{d})}(t-\tau) + C^{(\textrm{d})}(t-\nu).
\end{align*}  

Since $C^{(\textrm{d})}$ is non decreasing and $\tau > \nu$, we get $C^{(\textrm{d})}(t-\nu) > C^{(\textrm{d})}(t-\tau)$. Hence:
\begin{align*}
&C_t(P_t,Q^\pi_t,N^\pi_t) \\
&> P_t (\alpha|N^{\pi'}_t|+\beta) + \sum_{H_i\in (Q^{\pi'}_t \setminus N^{\pi'}_t)}C^{(\textrm{d})}(t-i) \\
&= C_t(P_t,Q^{\pi'}_t,N^{\pi'}_t).
\end{align*} 

Altogether, we got
\begin{align*}
&C(\pi) \\
&= \mathbb{E}\left[\sum_t \gamma^t C_t(P_t,Q^\pi_t,N^\pi_t)\right] \\
&> \mathbb{E}\left[\sum_t \gamma^t C_t(P_t,Q^{\pi'}_t,N^{\pi'}_t)\right] \\
&= C(\pi')
\end{align*}
and thus $\pi$ is not an optimal policy.
\qed\end{proof}

\subsection{Theorem \ref{thm:optimal_threshold}}
\begin{proof}
First, note that:
\begin{align*}
&\mathbb{E}\left[R^{[n]}(P)\right] \\
&= \mathbb{E}\left[P\prod_{i=1}^n{e^{\mathcal{N}(\mu,\sigma^2)}}\right] \\
&= P\prod_{i=1}^n{\mathbb{E}\left[e^{\mathcal{N}(\mu,\sigma^2)}\right]} \\
&= P e^{n(\mu+\frac{\sigma^2}{2})}.
\end{align*}

Substituting all parameters in Eq. \ref{eq:publish_threshold}, we get the following optimal criteria for publishing transaction $H_\tau$ at state $(t,P)$:
\begin{align*}
&0 \leq \alpha \left(\gamma^n e^{n(\mu+\frac{\sigma^2}{2})} - 1\right) P + 2c \sum_{i=0}^{n-1}{\gamma^i (t-\tau+i)} \\
&= \alpha \left(\gamma^n e^{n(\mu+\frac{\sigma^2}{2})} - 1\right) P + \frac{2c}{1-\gamma} \left( (1-\gamma^n)(t-\tau+\frac{\gamma}{1-\gamma}) - n \gamma^n \right),
\end{align*}
and thus,
\[ P  \leq \frac{2c}{\alpha(1-\gamma)} \frac{(1-\gamma^n)(t-\tau+\frac{\gamma}{1-\gamma}) - n \gamma^n}{1 - \gamma^n e^{n(\mu+\frac{\sigma^2}{2})}} \]
for all $n\geq 1$, as desired.
\qed\end{proof}

\subsection{Corollary \ref{cor:martingale}}
\begin{proof}
Substitute $\mu=-\frac{\sigma^2}{2}$ in Theorem \ref{thm:optimal_threshold} and get:
\begin{align*}
& \lambda(x) \\
&=\frac{2c}{1-\gamma} \inf_{n\geq 1} \frac{(1-\gamma^n)(x+\frac{\gamma}{1-\gamma})-n\gamma^n}{1-\gamma^n} \\
&= \frac{2c}{1-\gamma}\left(x+\frac{\gamma}{1-\gamma}-\sup_{n\geq 1}\frac{n\gamma^n}{1-\gamma^n} \right)
\end{align*}

If we show that $f(n) \triangleq \frac{n\gamma^n}{1-\gamma^n}$ is decreasing, then the maximum over $n\geq 1$ is obtained at the border $n=1$, and thus $\lambda(x) = \frac{2cx}{1-\gamma}$.

To show that $f(n)$ is decreasing, we will first find the derivative:
\[ f'(n) = \frac{\gamma^n \left(1 - \gamma^n + \ln(\gamma^n)\right)}{(1-\gamma^n)^2}. \]
Since $\ln(\gamma^n) \leq \gamma^n - 1$, we get that $f'(n) \leq 0$ and thus $f$ is decreasing as needed.
\qed\end{proof}

\end{document}